\documentclass[nofootinbib,prd,showpacs,showkeys]{revtex4}%
\usepackage{amssymb}
\usepackage{amsmath}
\usepackage{amsfonts}
\usepackage{graphicx}%
\begin{document}
\title{Modified Dispersion Relations: from Black-Hole Entropy to the Cosmological Constant.}
\author{Remo Garattini}
\email{remo.garattini@unibg.it.}
\affiliation{Universit\`{a} degli Studi di Bergamo, Facolt\`{a} di Ingegneria,}
\affiliation{Viale Marconi 5, 24044 Dalmine (Bergamo) Italy.}
\affiliation{I.N.F.N. - sezione di Milano, Milan, Italy.}

\begin{abstract}
Quantum Field Theory is plagued by divergences in the attempt to calculate
physical quantities. Standard techniques of regularization and renormalization
are used to keep under control such a problem. In this paper we would like to
use a different scheme based on Modified Dispersion Relations (MDR) to remove
infinities appearing in one loop approximation in contrast to what happens in
conventional approaches. In particular, we apply the MDR regularization to the
computation of the entropy of a Schwarzschild black hole from one side and the
Zero Point Energy (ZPE) of the graviton from the other side. The graviton ZPE
is connected to the cosmological constant by means of of the Wheeler-DeWitt equation.

\end{abstract}
\maketitle

\section{Introduction}

The appearance of a trans-Planckian physics in Black Hole thermodynamics has
led many authors to consider that some deep change in particle physics should
come into play. One realization of these ideas is represented by the
modification of the Heisenberg uncertainty relations, better known as
\textit{Generalized Uncertainty Principle} (GUP)\cite{Xiang
Li,RenQinChun,GAC1,Mod}. This principle is based on the following inequality%
\begin{equation}
\Delta x\Delta p\geq\hbar+\frac{\lambda_{p}^{2}}{\hbar}\left(  \Delta
p\right)  ^{2},
\end{equation}
where $\hbar$ is the Planck constant and $\lambda_{p}$ is the Planck length.
Of course, the above inequality affects the Liouville measure which becomes%
\begin{equation}
\frac{d^{3}xd^{3}p}{\left(  2\pi\hbar\right)  ^{3}\left(  1+\lambda
p^{2}\right)  ^{3}}. \label{eqn:states}%
\end{equation}
When $\lambda=0$, the formula reduces to the ordinary counting of quantum
states. If Eq.$\left(  \ref{eqn:states}\right)  $ is used for computing the
entropy of a black hole from a Quantum Field Theory point of view, the usual
UV divergence at the horizon can be removed\cite{Xiang Li,RenQinChun}. Indeed,
without introducing Eq.$\left(  \ref{states}\right)  $ one is forced to use
traditional methods for removing divergences: for example renormalizing the
Newton constant\cite{RenG}, or using Pauli-Villars regularization\cite{PV}. It
is clear that the distortion of the Liouville measure plays a key r\^{o}le in
regularizing divergent integrals. Non-commutative geometry provides another
powerful method to have such a distortion. As shown in
\cite{CPL,NCthermo,NCstat}, one finds\cite{RG PN}%
\begin{equation}
dn=\frac{d^{3}xd^{3}k}{\left(  2\pi\right)  ^{3}}\ \Longrightarrow
\ dn=\frac{d^{3}xd^{3}k}{\left(  2\pi\right)  ^{3}}\exp\left(  -\frac{\theta
}{4}k^{2}\right)  . \label{moddn}%
\end{equation}
This deformation corresponds to an effective cut off on the background
geometry. The UV cut off is triggered only by higher momenta modes
$\gtrsim1/\sqrt{\theta}$ which propagate over the background geometry. When
$\theta=0$, the formula reduces to the ordinary counting of quantum states. An
application of non-commutative geometry to the computation of black hole
entropy shows that the usual horizon divergence disappears\cite{BaiYan}. In
connection with these ideas, in recent years, there has been a proposal on how
the fundamental aspects of special relativity can be modified at very high
energies. This modification has been termed \textit{Doubly Special Relativity}
(DSR)\cite{GAC}. In DSR, the Planck mass is regarded as an observer
independent energy scale. One of its effects is that the usual dispersion
relation of a massive particle of mass $m$ is modified into the following
expression%
\begin{equation}
E^{2}g_{1}^{2}\left(  E/E_{P}\right)  -p^{2}g_{2}^{2}\left(  E/E_{P}\right)
=m^{2}, \label{mdisp}%
\end{equation}
where $g_{1}\left(  E/E_{P}\right)  $ and $g_{2}\left(  E/E_{P}\right)  $ are
two unknown functions which have the following property%
\begin{equation}
\lim_{E/E_{P}\rightarrow0}g_{1}\left(  E/E_{P}\right)  =1\qquad\text{and}%
\qquad\lim_{E/E_{P}\rightarrow0}g_{2}\left(  E/E_{P}\right)  =1. \label{prop}%
\end{equation}
Thus, the usual dispersion relation is recovered at low energies. Eqs.$\left(
\ref{mdisp},\ref{prop}\right)  $ are a representation of \textquotedblleft%
\textit{Modified Dispersion Relations\textquotedblright} (MDRs). The common
motivation in using them is in that they can be used as a phenomenological
approach to investigate physics at the Planck scale, where General Relativity
is no longer reliable. Moreover, we expect the functions $g_{1}\left(
E/E_{P}\right)  $ and $g_{2}\left(  E/E_{P}\right)  $ modify the UV behavior
of quantum fields in the same way as GUP and Non-commutative geometry do,
respectively. Note that GUP and MDR modifications are strictly
connected\cite{CG}. Since the form of $g_{1}\left(  E/E_{P}\right)  $ and
$g_{2}\left(  E/E_{P}\right)  $ is unknown and they have to obey the property
$\left(  \ref{prop}\right)  $, we have a large amount of arbitrariness in
fixing the dependence on $E/E_{P}$, even if some specific choices have been
proposed by G. Amelino-Camelia et al.\cite{Amelino et Al.} in the context of
black hole thermodynamics. MDRs play a relevant r\^{o}le also when the
background is curved. Following the analysis of Magueijo and
Smolin\cite{MagSmo} one can define the following \textquotedblleft%
\textit{rainbow metric}\textquotedblright%
\begin{equation}
ds^{2}=-\frac{N^{2}\left(  r\right)  dt^{2}}{g_{1}^{2}\left(  E/E_{P}\right)
}+\frac{dr^{2}}{\left(  1-\frac{b\left(  r\right)  }{r}\right)  g_{2}%
^{2}\left(  E/E_{P}\right)  }+\frac{r^{2}}{g_{2}^{2}\left(  E/E_{P}\right)
}\left(  d\theta^{2}+\sin^{2}\theta d\phi^{2}\right)  \label{line}%
\end{equation}
which is a solution of the distorted Einstein's Field equations%
\begin{equation}
G_{\mu\nu}\left(  E\right)  =8\pi G\left(  E\right)  T_{\mu\nu}\left(
E\right)  +g_{\mu\nu}\Lambda\left(  E\right)  .
\end{equation}
$G\left(  E\right)  $ is an energy dependent Newton's constant, defined so
that $G\left(  0\right)  $ is the physical Newton's constant. Similarly we
have an energy dependent cosmological constant $\Lambda\left(  E\right)  $.
The function $b\left(  r\right)  $ will be referred to as the
\textquotedblleft shape function\textquotedblright. The shape function may be
thought of as specifying the shape of the spatial slices. If the equation
$b\left(  r_{w}\right)  =r_{w}$ is satisfied for some values of $r$, then we
say that the points $r_{w}$ are horizons for the metric $\left(
\ref{line}\right)  $. In this paper we will discuss the use of MDRs on black
hole entropy calculation\cite{RemoPLB} and on the estimation of the
cosmological constant computed with the help of a revisited Wheeler-DeWitt
equation (WDW)\cite{RGGM}. Units in which $\hbar=c=k=1$ are used throughout
the paper.

\section{Black Hole Entropy with MDRs}

\label{p2}To start with, we fix the ideas on a real massless scalar field
described by the action
\begin{equation}
I=-\frac{1}{2}\int d^{4}x\sqrt{-g}\left[  g^{\mu\nu}\partial_{\mu}\phi
\partial_{\nu}\phi\right]
\end{equation}
in the background geometry of Eq.$\left(  \ref{line}\right)  $ with $N\left(
r\right)  $ described by%
\begin{equation}
N^{2}\left(  r\right)  =\exp\left(  -2A\left(  r\right)  \right)  \left(
1-\frac{b\left(  r\right)  }{r}\right)  , \label{N(r)}%
\end{equation}
where $A\left(  r\right)  $ is known as the \textquotedblleft redshift
function\textquotedblright\ that describes how far the total gravitational
redshift deviates from that implied by the shape function. Without loss of
generality we can fix the value of $A\left(  r\right)  $ at infinity such that
$A\left(  \infty\right)  =0$. The Euler-Lagrange equations are%
\begin{equation}
\frac{1}{\sqrt{-g}}\partial_{\mu}\left(  \sqrt{-g}g^{\mu\nu}\partial_{\nu
}\right)  \phi=0.
\end{equation}
In order to use the WKB approximation, if $\phi$ has a separable form, we can
define an r-dependent radial wave number $k(r,l,E)$
\begin{equation}
k_{r}^{2}(r,l,E)\equiv\frac{1}{\left(  1-\frac{b\left(  r\right)  }{r}\right)
}\left[  \exp\left(  2A\left(  r\right)  \right)  \frac{E^{2}h^{2}\left(
E/E_{P}\right)  }{\left(  1-\frac{b\left(  r\right)  }{r}\right)  }%
-\frac{l(l+1)}{r^{2}}\right]  , \label{squarekr}%
\end{equation}
with%
\begin{equation}
h\left(  E/E_{P}\right)  =\frac{g_{1}\left(  E/E_{P}\right)  }{g_{2}\left(
E/E_{P}\right)  }. \label{h(E)}%
\end{equation}
The number of modes with frequency less than $E$ is given approximately by
\begin{equation}
\tilde{g}(E)=\frac{1}{\pi}\int_{0}^{l_{max}}(2l+1)\int_{r_{w}}^{R}\sqrt
{k^{2}(r,l,E)}drdl,
\end{equation}
where it is understood that the integration with respect to $r$ and $l$ is
taken over those values which satisfy $r_{w}\leq r\leq R$ and $k^{2}%
(r,l,E)\geq0$. Thus, from Eq.$\left(  \ref{squarekr}\right)  $ we get%
\begin{equation}
\frac{d\tilde{g}(E)}{dE}=\int\frac{\partial\nu(l{,E})}{\partial E}%
(2l+1)dl=\frac{2}{\pi}\frac{d}{dE}\left(  \frac{1}{3}E^{3}h^{3}\left(
E\right)  \right)  \int_{r_{w}}^{R}dr\frac{\exp\left(  3A\left(  r\right)
\right)  }{\left(  1-\frac{b\left(  r\right)  }{r}\right)  ^{2}}r^{2}.
\label{states}%
\end{equation}
Defining $\beta$ as the inverse temperature measured at infinity, the free
energy is given by
\begin{equation}
F=\frac{1}{\beta}\int_{0}^{\infty}\ln\left(  1-e^{-\beta E}\right)
\frac{d\tilde{g}(E)}{dE}dE=F_{r_{w}}+F_{R}.
\end{equation}
We divide the integration range into two intervals: in $\left[  r_{w}%
,r_{1}\right]  $ we define $F_{r_{w}}$and in $\left[  r_{1},+\infty\right)  $
with $r_{1}>r_{w}$ we define $F_{R}$. Assuming that $A\left(  r\right)
<\infty,$ $\forall r\in\left[  r_{w},+\infty\right)  $, $F_{R}$ is dominated
by large volume effects for large $R$. This term will give the contribution to
the entropy of a homogeneous quantum gas in flat space at a uniform
temperature $T$. We fix our attention on%
\begin{equation}
F_{r_{w}}=\frac{2}{\pi}\frac{1}{\beta}\int_{0}^{\infty}\ln\left(  1-e^{-\beta
E}\right)  \frac{d}{dE}\left(  \frac{1}{3}E^{3}h^{3}\left(  E\right)  \right)
dE\int_{r_{w}}^{r_{1}}drr^{2}\frac{\exp\left(  3A\left(  r\right)  \right)
}{\left(  1-\frac{b\left(  r\right)  }{r}\right)  ^{2}}. \label{Frw0}%
\end{equation}
In proximity of $r_{w}$%
\begin{equation}
1-\frac{b\left(  r\right)  }{r}=\frac{r-r_{w}}{r_{w}}\text{ }\left[
1-b^{\prime}\left(  r_{w}\right)  \right]  \label{e32a}%
\end{equation}
and the radial part of $F_{r_{w}}$ becomes divergent. This ultraviolet
divergence has been cured by 't Hooft, who introduced a \textquotedblleft%
\textit{brick wall }$r_{0}$\textquotedblright\ proportional to $l_{P}^{2}%
$\cite{tHooft}. Nevertheless, since spacetime is modified by a
\textquotedblleft\textit{rainbow metric}\textquotedblright, it is quite
natural that even the \textquotedblleft\textit{brick wall}\textquotedblright%
\ is affected by this distortion. To see such an effect, we perform the radial
integration in $F_{r_{w}}$ to obtain%
\begin{equation}
\int_{r_{w}+r_{0}}^{r_{1}}drr^{2}\frac{\exp\left(  3A\left(  r\right)
\right)  }{\left(  1-\frac{b\left(  r\right)  }{r}\right)  ^{2}}=\int
_{r_{w}+r\left(  E/E_{P}\right)  }^{r_{1}}drr^{2}\frac{\exp\left(  3A\left(
r\right)  \right)  }{\left(  1-\frac{b\left(  r\right)  }{r}\right)  ^{2}%
}\simeq r_{w}^{4}\frac{\exp\left(  3A\left(  r_{w}\right)  \right)  }{\left(
1-b^{\prime}\left(  r_{w}\right)  \right)  ^{2}}\frac{1}{r\left(
E/E_{P}\right)  }, \label{rw}%
\end{equation}
where we have assumed that, in proximity of the throat the brick wall is no
longer a constant but it becomes a function of $E/E_{P}$. Plugging Eq.$\left(
\ref{rw}\right)  $ into Eq.$\left(  \ref{Frw0}\right)  $ we find%
\begin{equation}
F_{r_{w}}=-\frac{2r_{w}^{4}}{3\beta\pi}\frac{\exp\left(  3\Lambda\left(
r_{w}\right)  \right)  }{\left(  1-b^{\prime}\left(  r_{w}\right)  \right)
^{2}}\int_{0}^{\infty}E^{3}h^{3}\left(  E/E_{P}\right)  \frac{d}{dE}\left[
\frac{\ln\left(  1-\exp\left(  -\beta E\right)  \right)  }{r\left(
E/E_{P}\right)  }\right]  dE, \label{Frw}%
\end{equation}
where we have integrated by parts with the condition that $h\left(
E/E_{P}\right)  $ be chosen in such a way to allow the convergence when
$E/E_{P}\rightarrow\infty$. Without loss of generality we write%
\begin{equation}
r\left(  E/E_{P}\right)  =r_{w}\sigma\left(  E/E_{P}\right)  , \label{r(E)}%
\end{equation}
with%
\begin{equation}
\sigma\left(  E/E_{P}\right)  \rightarrow0,\qquad E/E_{P}\rightarrow0.
\label{r(E)sigma}%
\end{equation}
In this way the horizon divergence is still present but it is translated in
terms of a function of $E/E_{P}$. Plugging Eq.$\left(  \ref{r(E)}\right)  $
into Eq.$\left(  \ref{Frw}\right)  $, we obtain%
\[
F_{r_{w}}=-\frac{C_{r_{w}}}{3\beta r_{w}}\int_{0}^{\infty}E^{3}h^{3}\left(
E/E_{P}\right)  \frac{d}{dE}\left[  \frac{\ln\left(  1-\exp\left(  -\beta
E\right)  \right)  }{\sigma\left(  E/E_{P}\right)  }\right]  dE
\]
where we have defined%
\begin{equation}
C_{r_{w}}=\frac{2r_{w}^{4}}{\pi}\frac{\exp\left(  3\Lambda\left(
r_{w}\right)  \right)  }{\left(  1-b^{\prime}\left(  r_{w}\right)  \right)
^{2}}. \label{e33a}%
\end{equation}
It is clear that $g_{1}\left(  E/E_{P}\right)  $ and $g_{2}\left(
E/E_{P}\right)  $ must be chosen in such a way to compensate the vanishing of
$\sigma\left(  E/E_{P}\right)  $, otherwise the horizon divergence
(\textit{brick wall}) cannot be eliminated. For example, one good candidate
for the convergence is%
\begin{equation}
h\left(  E/E_{P}\right)  =\exp\left(  -\frac{E}{E_{P}}\right)  . \label{hexp}%
\end{equation}
A good candidate, but not exhaustive for $\sigma\left(  E/E_{P}\right)  $ is
\begin{equation}
\sigma\left(  E/E_{P}\right)  =\left(  \frac{E}{E_{P}}\right)  ^{\alpha
},\qquad\alpha>0. \label{sigma(E)}%
\end{equation}
In the limit where $\beta E_{P}\gg1$, the total energy $U$ is%
\begin{equation}
U=\frac{\partial\left(  \beta F_{r_{w}}\right)  }{\partial\beta}=r_{w}%
^{2}\frac{\exp\left(  2A\left(  r_{w}\right)  \right)  }{1-b^{\prime}\left(
r_{w}\right)  }\frac{2E_{P}^{2}}{9\beta^{2}\kappa_{w}}\pi=r_{w}^{2}\frac
{\exp\left(  2A\left(  r_{w}\right)  \right)  }{1-b^{\prime}\left(
r_{w}\right)  }\frac{E_{P}^{2}}{9\beta}%
\end{equation}
and the entropy $S$ is%
\begin{equation}
S=\beta^{2}\frac{\partial F_{r_{w}}}{\partial\beta}=r_{w}^{2}\frac{\exp\left(
2\Lambda\left(  r_{w}\right)  \right)  }{1-b^{\prime}\left(  r_{w}\right)
}\frac{4E_{p}^{2}}{9\beta\kappa_{w}}\pi=\frac{A_{r_{w}}E_{P}^{2}}{4}\frac
{\exp\left(  2\Lambda\left(  r_{w}\right)  \right)  }{1-b^{\prime}\left(
r_{w}\right)  }\frac{2}{9\pi},
\end{equation}
where we have used the expression for the surface gravity in the low energy
limit%
\begin{equation}
\kappa_{w}=\frac{1}{2r_{w}}\exp\left(  -A\left(  r_{w}\right)  \right)
\left[  1-b^{\prime}\left(  r_{w}\right)  \right]  \label{kwl}%
\end{equation}
and where we have integrated over $E$. To recover the area law, we have to
impose that%
\begin{equation}
\frac{\exp\left(  2\Lambda\left(  r_{w}\right)  \right)  }{1-b^{\prime}\left(
r_{w}\right)  }=\frac{9\pi}{2}%
\end{equation}
and%
\begin{equation}
\frac{1}{\beta}=T=\frac{\kappa_{w}}{2\pi}. \label{e35}%
\end{equation}
This corresponds to a changing of the time variable with respect to the
Schwarzschild time. The total energy becomes%
\begin{equation}
U=r_{w}^{2}\frac{\pi E_{P}^{2}}{2\beta}, \label{e36}%
\end{equation}
which in terms of the Schwarzschild radius $r_{w}=2MG$ and inverse Hawking
temperature $\beta=8\pi MG$ becomes%
\begin{equation}
U=4M^{2}G^{2}\frac{E_{P}^{2}}{16MG}=\frac{M}{4}.
\end{equation}
Note the discrepancy of a factor of $3/2$ with the 't Hooft result.

\section{Gravity's Rainbow and the WDW Equation}

The WDW equation was originally introduced by Bryce DeWitt as an attempt to
quantize General Relativity in a Hamiltonian formulation. It is described
by\cite{DeWitt}%
\begin{equation}
\mathcal{H}\Psi=\left[  \left(  2\kappa\right)  G_{ijkl}\pi^{ij}\pi^{kl}%
-\frac{\sqrt{g}}{2\kappa}\!{}\!\left(  \,\!^{3}R-2\Lambda\right)  \right]
\Psi=0 \label{WDW}%
\end{equation}
and it represents the quantum version of the classical constraint which
guarantees the invariance under time reparametrization. $G_{ijkl}$ is the
super-metric, $\pi^{ij}$ is the super-momentum,$^{3}R$ is the scalar curvature
in three dimensions and $\Lambda$ is the cosmological constant, while
$\kappa=8\pi G$ with $G$ the Newton's constant. In this way, the WDW equation
is written in its most general form. The main reason to use such an equation
to discuss renormalization problems is related to the possibility of formally
re-writing the WDW equation as an expectation value computation. Rather than
reproduce the formalism, we shall refer the reader to Refs.\cite{Remo} for
details, when necessary. However, for self-completeness and self-consistency,
we present here a brief outline of the formalism used\footnote{See also
Ref.\cite{CG:2007} for an application of the method to a $f\left(  R\right)  $
theory.}. Multiplying Eq.$\left(  \ref{WDW}\right)  $ by $\Psi^{\ast}\left[
g_{ij}\right]  $ and functionally integrating over the three spatial metric
$g_{ij}$ we find%
\begin{equation}
\frac{1}{V}\frac{\int\mathcal{D}\left[  g_{ij}\right]  \Psi^{\ast}\left[
g_{ij}\right]  \int_{\Sigma}d^{3}x\hat{\Lambda}_{\Sigma}\Psi\left[
g_{ij}\right]  }{\int\mathcal{D}\left[  g_{ij}\right]  \Psi^{\ast}\left[
g_{ij}\right]  \Psi\left[  g_{ij}\right]  }=\frac{1}{V}\frac{\left\langle
\Psi\left\vert \int_{\Sigma}d^{3}x\hat{\Lambda}_{\Sigma}\right\vert
\Psi\right\rangle }{\left\langle \Psi|\Psi\right\rangle }=-\frac{\Lambda
}{\kappa}. \label{VEV}%
\end{equation}
In Eq.$\left(  \ref{VEV}\right)  $ we have also integrated over the
hypersurface $\Sigma$ and we have defined%
\begin{equation}
V=\int_{\Sigma}d^{3}x\sqrt{g}.
\end{equation}
$V$ is the volume of the hypersurface $\Sigma$ and%
\begin{equation}
\hat{\Lambda}_{\Sigma}=\left(  2\kappa\right)  G_{ijkl}\pi^{ij}\pi^{kl}%
-\sqrt{g}^{3}R/\left(  2\kappa\right)  . \label{LambdaSigma}%
\end{equation}
In this form, Eq.$\left(  \ref{VEV}\right)  $ can be used to compute ZPE
provided that $\Lambda/\kappa$ be considered as an eigenvalue of $\hat
{\Lambda}_{\Sigma}$. In particular, Eq.$\left(  \ref{VEV}\right)  $ represents
the Sturm-Liouville problem associated with the cosmological constant. To
solve Eq.$\left(  \ref{VEV}\right)  $ is a quite impossible task. Therefore,
we are oriented to use a variational approach with trial wave functionals. The
related boundary conditions are dictated by the choice of the trial wave
functionals which, in our case are of the Gaussian type. Different types of
wave functionals correspond to different boundary conditions. The choice of a
Gaussian wave functional is justified by the fact that ZPE should be described
by a good candidate of the \textquotedblleft\textit{vacuum state}%
\textquotedblright. To fix ideas, we choose the line element $\left(
\ref{line}\right)  $ as background metric with $g_{1}\left(  E/E_{P}\right)
=g_{2}\left(  E/E_{P}\right)  =1$, namely MDRs do not distort the metric. Then
we consider a perturbation of the metric tensor of the form $g_{ij}=\bar
{g}_{ij}+h_{ij}$, where $\bar{g}_{ij}$ is the background metric and $h_{ij}$
is a quantum fluctuation around the background. Thus Eq.$\left(
\ref{VEV}\right)  $ can be expanded in terms of $h_{ij}$. Since the kinetic
part of $\hat{\Lambda}_{\Sigma}$ is quadratic in the momenta, we only need to
expand the three-scalar curvature $\int d^{3}x\sqrt{g}{}^{3}R$ up to the
quadratic order. To proceed with the computation, we need an orthogonal
decomposition on the tangent space of 3-metric deformations\cite{Quad}:%

\begin{equation}
h_{ij}=\frac{1}{3}\left(  \sigma+2\nabla\cdot\xi\right)  g_{ij}+\left(
L\xi\right)  _{ij}+h_{ij}^{\bot}. \label{p21a}%
\end{equation}
The operator $L$ maps $\xi_{i}$ into symmetric tracefree tensors%
\begin{equation}
\left(  L\xi\right)  _{ij}=\nabla_{i}\xi_{j}+\nabla_{j}\xi_{i}-\frac{2}%
{3}g_{ij}\left(  \nabla\cdot\xi\right)  ,
\end{equation}
$h_{ij}^{\bot}$ is the traceless-transverse component of the perturbation
(TT), namely $g^{ij}h_{ij}^{\bot}=0$, $\nabla^{i}h_{ij}^{\bot}=0$ and $h$ is
the trace of $h_{ij}$. It is immediate to recognize that the trace element
$\sigma=h-2\left(  \nabla\cdot\xi\right)  $ is gauge invariant. The same
decomposition can be done also on the momentum $\pi^{ij}$ and induces the
following transformation on the functional measure $\mathcal{D}h_{ij}%
\rightarrow\mathcal{D}h_{ij}^{\bot}\mathcal{D}\xi_{i}\mathcal{D}\sigma J_{1}$,
where $J_{1}$ is the Jacobian related to the gauge vector variable $\xi_{i}$.
The only physical information is encoded%
\begin{equation}
\frac{1}{V}\frac{\left\langle \Psi^{\bot}\left\vert \int_{\Sigma}d^{3}x\left[
\hat{\Lambda}_{\Sigma}^{\bot}\right]  ^{\left(  2\right)  }\right\vert
\Psi^{\bot}\right\rangle }{\left\langle \Psi^{\bot}|\Psi^{\bot}\right\rangle
}=-\frac{\Lambda^{\bot}}{\kappa}. \label{lambda0_2a}%
\end{equation}
After having functionally integrated, we find%
\begin{equation}
\hat{\Lambda}_{\Sigma}^{\bot}=\frac{1}{4V}\int_{\Sigma}d^{3}x\sqrt{\bar{g}%
}G^{ijkl}\left[  \left(  2\kappa\right)  K^{-1\bot}\left(  x,x\right)
_{ijkl}+\frac{1}{\left(  2\kappa\right)  }\!{}\left(  \tilde{\bigtriangleup
}_{L\!}\right)  _{j}^{a}K^{\bot}\left(  x,x\right)  _{iakl}\right]  ,
\label{p22}%
\end{equation}
where%
\begin{equation}
\left(  \tilde{\bigtriangleup}_{L\!}\!{}h^{\bot}\right)  _{ij}=\left(
\bigtriangleup_{L\!}\!{}h^{\bot}\right)  _{ij}-4R{}_{i}^{k}\!{}h_{kj}^{\bot
}+\text{ }^{3}R{}\!{}h_{ij}^{\bot} \label{M Lichn}%
\end{equation}
is the modified Lichnerowicz operator and $\bigtriangleup_{L}$is the
Lichnerowicz operator defined by%
\begin{equation}
\left(  \bigtriangleup_{L}h\right)  _{ij}=\bigtriangleup h_{ij}-2R_{ikjl}%
h^{kl}+R_{ik}h_{j}^{k}+R_{jk}h_{i}^{k}\qquad\bigtriangleup=-\nabla^{a}%
\nabla_{a}. \label{DeltaL}%
\end{equation}
$G^{ijkl}$ represents the inverse DeWitt metric and all indices run from one
to three. Note that the term $-4R{}_{i}^{k}\!{}h_{kj}^{\bot}+$ $^{3}R{}%
\!{}h_{ij}^{\bot}$ disappears in four dimensions. The propagator $K^{\bot
}\left(  x,x\right)  _{iakl}$ can be represented as
\begin{equation}
K^{\bot}\left(  \overrightarrow{x},\overrightarrow{y}\right)  _{iakl}%
=\sum_{\tau}\frac{h_{ia}^{\left(  \tau\right)  \bot}\left(  \overrightarrow
{x}\right)  h_{kl}^{\left(  \tau\right)  \bot}\left(  \overrightarrow
{y}\right)  }{2\lambda\left(  \tau\right)  }, \label{proptt}%
\end{equation}
where $h_{ia}^{\left(  \tau\right)  \bot}\left(  \overrightarrow{x}\right)  $
are the eigenfunctions of $\tilde{\bigtriangleup}_{L\!}$. $\tau$ denotes a
complete set of indices and $\lambda\left(  \tau\right)  $ are a set of
variational parameters to be determined by the minimization of Eq.$\left(
\ref{p22}\right)  $. The expectation value of $\hat{\Lambda}_{\Sigma}^{\bot}$
is easily obtained by inserting the form of the propagator into Eq.$\left(
\ref{p22}\right)  $ and minimizing with respect to the variational function
$\lambda\left(  \tau\right)  $. Thus the total one loop energy density for TT
tensors becomes%
\begin{equation}
\frac{\Lambda}{8\pi G}=-\frac{1}{2}\sum_{\tau}\left[  \sqrt{\omega_{1}%
^{2}\left(  \tau\right)  }+\sqrt{\omega_{2}^{2}\left(  \tau\right)  }\right]
. \label{1loop}%
\end{equation}
The above expression makes sense only for $\omega_{i}^{2}\left(  \tau\right)
>0$, where $\omega_{i}$ are the eigenvalues of $\tilde{\bigtriangleup}_{L\!}$.
For a background of the form of Eq.$\left(  \ref{line}\right)  $, if we define
the reduced fields $f_{i}\left(  x\right)  =F_{i}\left(  x\right)  /r$, we
find that the Lichnerowicz operator $\left(  \tilde{\bigtriangleup}_{L\!}%
\!{}h^{\bot}\right)  _{ij}$ can be reduced to%
\begin{equation}
\left[  -\frac{d^{2}}{dx^{2}}+\frac{l\left(  l+1\right)  }{r^{2}}+m_{i}%
^{2}\left(  r\right)  \right]  f_{i}\left(  x\right)  =\omega_{i,l}^{2}%
f_{i}\left(  x\right)  \quad i=1,2\quad, \label{p34}%
\end{equation}
where we have used the Regge-Wheeler representation\cite{Regge Wheeler:1957}
and $m_{1}^{2}\left(  r\right)  $ and $m_{2}^{2}\left(  r\right)  $ are two
r-dependent effective masses%
\begin{equation}
\left\{
\begin{array}
[c]{c}%
m_{1}^{2}\left(  r\right)  =\frac{6}{r^{2}}\left(  1-\frac{b\left(  r\right)
}{r}\right)  +\frac{3}{2r^{2}}b^{\prime}\left(  r\right)  -\frac{3}{2r^{3}%
}b\left(  r\right) \\
\\
m_{2}^{2}\left(  r\right)  =\frac{6}{r^{2}}\left(  1-\frac{b\left(  r\right)
}{r}\right)  +\frac{1}{2r^{2}}b^{\prime}\left(  r\right)  +\frac{3}{2r^{3}%
}b\left(  r\right)
\end{array}
\right.  \quad\left(  r\equiv r\left(  x\right)  \right)  . \label{masses}%
\end{equation}
In Eq.$\left(  \ref{p34}\right)  $ , we have defined%
\begin{equation}
dx=\pm\frac{dr}{\sqrt{1-\frac{b\left(  r\right)  }{r}}}. \label{dx}%
\end{equation}
Like in the entropy calculation of section $\left(  \ref{p2}\right)  $, we use
the W.K.B. method and we define two r-dependent radial wave numbers%
\begin{equation}
k_{i}^{2}\left(  r,l,\omega_{i,nl}\right)  =\omega_{i,nl}^{2}-\frac{l\left(
l+1\right)  }{r^{2}}-m_{i}^{2}\left(  r\right)  \quad i=1,2\quad. \label{kTT}%
\end{equation}
For every degree of freedom of the graviton we apply Eq.$\left(
\ref{states}\right)  $ and we find that Eq.$\left(  \ref{1loop}\right)  $
becomes%
\begin{equation}
\frac{\Lambda}{8\pi G}=-\frac{1}{\pi}\sum_{i=1}^{2}\int_{0}^{+\infty}%
\omega_{i}\frac{d\tilde{g}\left(  \omega_{i}\right)  }{d\omega_{i}}d\omega
_{i}. \label{tot1loop}%
\end{equation}
This is the one loop graviton contribution to the induced cosmological
constant. The explicit evaluation of Eq.$\left(  \ref{tot1loop}\right)  $
gives%
\begin{equation}
\frac{\Lambda}{8\pi G}=\rho_{1}+\rho_{2}=-\frac{1}{4\pi^{2}}\sum_{i=1}^{2}%
\int_{\sqrt{m_{i}^{2}\left(  r\right)  }}^{+\infty}\omega_{i}^{2}\sqrt
{\omega_{i}^{2}-m_{i}^{2}\left(  r\right)  }d\omega_{i}, \label{t1l}%
\end{equation}
where we have included an additional $4\pi$ coming from the angular
integration. $\rho_{1}$ and $\rho_{2}$ are divergent and traditionally the use
of the zeta function regularization keeps the divergences under control. To
this purpose we reconsider Eq.$\left(  \ref{WDW}\right)  $ in presence of
Gravity's Rainbow and we find\footnote{Details of the calculation in presence
of Gravity's Rainbow can be found in Ref.\cite{RGGM}}%
\begin{equation}
\frac{g_{2}^{3}\left(  E/E_{P}\right)  }{\tilde{V}}\frac{\left\langle
\Psi\left\vert \int_{\Sigma}d^{3}x\tilde{\Lambda}_{\Sigma}\right\vert
\Psi\right\rangle }{\left\langle \Psi|\Psi\right\rangle }=-\frac{\Lambda
}{\kappa}, \label{WDW3}%
\end{equation}
where%
\begin{equation}
\tilde{\Lambda}_{\Sigma}=\left(  2\kappa\right)  \frac{g_{1}^{2}\left(
E/E_{P}\right)  }{g_{2}^{3}\left(  E/E_{P}\right)  }\tilde{G}_{ijkl}\tilde
{\pi}^{ij}\tilde{\pi}^{kl}\mathcal{-}\frac{\sqrt{\tilde{g}}\tilde{R}}{\left(
2\kappa\right)  g_{2}\left(  E/E_{P}\right)  }\!{}\!. \label{LambdaR}%
\end{equation}
The symbol \textquotedblleft$\sim$\textquotedblright\ indicates the quantity
computed in absence of rainbow's functions $g_{1}\left(  E/E_{P}\right)  $ and
$g_{2}\left(  E/E_{P}\right)  $. Of course, Eqs.$\left(  \ref{WDW3}\right)  $
and $\left(  \ref{LambdaR}\right)  $ reduce to the ordinary Eqs.$\left(
\ref{WDW},\ref{VEV}\right)  $ and $\left(  \ref{LambdaSigma}\right)  $ when
$E/E_{Pl}\rightarrow0$. By repeating the procedure leading to Eq.$\left(
\ref{p22}\right)  $, we find that the total one loop energy density becomes%
\begin{equation}
\frac{\Lambda}{8\pi G}=-\frac{1}{3\pi^{2}}\sum_{i=1}^{2}\int_{E^{\ast}%
}^{+\infty}E_{i}g_{1}\left(  E/E_{P}\right)  g_{2}\left(  E/E_{P}\right)
\frac{d}{dE_{i}}\sqrt{\left(  \frac{E_{i}^{2}}{g_{2}^{2}\left(  E/E_{P}%
\right)  }-m_{i}^{2}\left(  r\right)  \right)  ^{3}}dE_{i}, \label{LoverG}%
\end{equation}
where $E^{\ast}$ is the value which annihilates the argument of the root. To
further proceed, we choose a form of $g_{1}\left(  E/E_{P}\right)  $ and
$g_{2}\left(  E/E_{P}\right)  $ suggested by a Noncommutative geometry
analysis\cite{RG PN}. If we fix%
\begin{equation}
g_{1}\left(  E/E_{P}\right)  =\left(  1+\beta\frac{E}{E_{P}}\right)
\exp(-\alpha\frac{E^{2}}{E_{P}^{2}})\qquad\text{and}\qquad g_{2}\left(
E/E_{P}\right)  =1, \label{g1g22}%
\end{equation}
with $\alpha>0$ and $\beta\in%
%TCIMACRO{\U{211d} }%
%BeginExpansion
\mathbb{R}
%EndExpansion
$, Eq.$\left(  \ref{LoverG}\right)  $ can be easily integrated. However, it is
more useful to give the asymptotic expansion for large and small $x$, where
$x=\sqrt{m_{1,2}^{2}\left(  r\right)  /E_{P}^{2}}$. The asymptotic expansion
for large $x$ is%
\begin{equation}
\frac{\Lambda}{8\pi G}\simeq-{\frac{\left(  2\beta{\alpha}^{3/2}+\sqrt{\pi
}{\alpha}^{2}\right)  x}{4{\alpha}^{7/2}}-\frac{8\beta{\alpha}^{5/2}%
+3\sqrt{\pi}{\alpha}^{3}}{16{\alpha}^{11/2}x}+\frac{3}{128}}\,{\frac
{16\beta{\alpha}^{7/2}+5\sqrt{\pi}{\alpha}^{4}}{{\alpha}^{15/2}{x}^{3}}%
}+O\left(  x^{-4}\right)  , \label{AsL}%
\end{equation}
while for small $x$, one gets%
\begin{equation}
\frac{\Lambda}{8\pi G}\simeq-{\frac{4{\alpha}^{5/2}+3\sqrt{\pi}\beta{\alpha
}^{2}}{4{\alpha}^{9/2}}}+O\left(  x^{3}\right)  . \label{SmL}%
\end{equation}
If we set%
\begin{equation}
\beta=-{\frac{\sqrt{\alpha\pi}}{2}}, \label{as}%
\end{equation}
then the linear divergent term of the asymptotic expansion $\left(
\ref{AsL}\right)  $ disappears. This means that $\Lambda/8\pi G$ vanishes for
large $x$. On the other hand for small $x$ we get%
\begin{equation}
\frac{\Lambda}{8\pi G}\simeq{\frac{{3\pi-8}}{8{\alpha}^{2}}}+O\left(
x^{3}\right)  , \label{AsLSm}%
\end{equation}
where we have used the result of expansion $\left(  \ref{SmL}\right)  $. It is
possible to show that with choice $\left(  \ref{as}\right)  $, the induced
cosmological constant is always positive. Note that when $\beta=0$, the pure
\textquotedblleft\textit{Gaussian}\textquotedblright\ choice can not give a
positive induced cosmological constant\cite{RGGM}.

\end{document}